\documentclass[twocolumn,letter]{jpsj2}

\title{
Dynamic Nuclear Polarization in a Quantum Hall Corbino Disk
}

\author{
	Minoru \textsc{Kawamura}$^{1}$, 
	Hiroyuki \textsc{Takahashi}$^{1}$, 
	Satoru \textsc{Masubuchi}$^{1}$,
	Yoshiaki \textsc{Hashimoto}$^{2}$,
	Shingo \textsc{Katsumoto}$^{2,3}$,
	Kohei \textsc{Hamaya}$^{1,3}$
	and Tomoki \textsc{Machida}$^{1,3}$
}

\inst{
	$^{1}$Institute of Industrial Science, University of Tokyo,
		4-6-1 Komaba, Meguro-ku, Tokyo 153-8505, Japan \\
	$^{2}$Institute for Solid State Physics, University of Tokyo,
 		5-1-5 Kashiwanoha, Kashiwa 277-8581, Japan \\
	$^{3}$Institute for Nano Quantum Information Electronics, University of Tokyo,
		4-6-1 Komaba, Meguro-ku, Tokyo 153-8505, Japan \\
}

\abst{
Electrical polarization of nuclear spins 
is studied in a Corbino disk under a breakdown regime of the quantum Hall effect (QHE).
Since the edge channels are completely absent in the Corbino disk,
we conclude that the electric current flowing in  the bulk channel of a quantum Hall conductor
is relevant to dynamic nuclear polarization (DNP).
A pump and probe measurement demonstrates
that DNP emerges
near the critical voltage of the QHE breakdown.
The agreement of the onset voltage of DNP with that of the QHE breakdown
indicates that the underlying  origin of  DNP is closely related to 
that of the QHE breakdown.
}

\kword{quantum Hall effect breakdown, dynamic nuclear polarization, hyperfine interaction}

\begin{document}
\maketitle

\begin{figure}[b]
	\begin{center}
		\includegraphics[width=5cm]{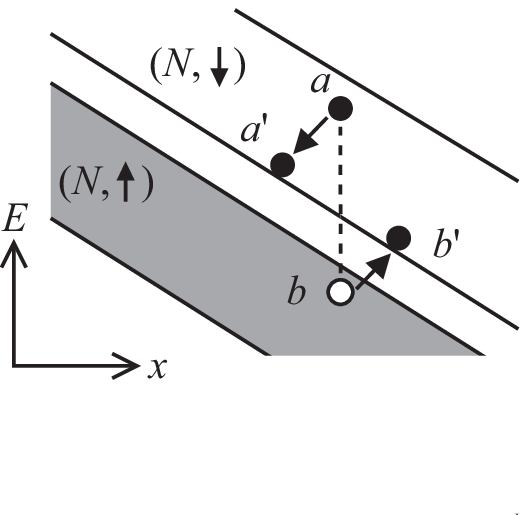}
		\caption{
		\label{impactionization}
		Schematic  of the impact ionization process. 
		The Landau subbands are tilted by the applied Hall electric field.
		An electron ({\it a}) in the upper Landau subband  collides with another electron 
		({\it b})
		in the lower subband  and  excites it 
		to the upper subband ({\it b}$\rightarrow${\it b}$^{\prime}$).
		}
	\end{center}
	\end{figure}

Breakdown of the quantum Hall effect (QHE) has been a subject of research 
interest not only for developing a QHE resistance standard
but also for understanding fundamental physics
of electron transport in quantum Hall (QH) 
systems\cite{Ebert1983,Cage1983,Dolgopolov1992, Nachtwei1999,Komiyama1996,Komiyama2000}.
When the electric field applied to  a two-dimensional electron gas (2DEG)
exceeds a critical value $E_{\rm c}$,
the QH conductor becomes unstable against the excitation 
of electron-hole ($e$-$h$) pairs,
leading to avalanche-type multiplication of the $e$-$h$ pairs\cite{Komiyama1996,Komiyama2000}.
As a result, the dissipationless QH state
breaks down and the longitudinal resistivity increases abruptly.
According to the earlier works on QHE breakdown\cite{Komiyama1996,Komiyama2000},
the multiplication of $e$-$h$ pairs occurs 
due to inter-Landau-subband impact ionization;
the electrons accelerated in the upper Landau subband
collide with the other electrons in the lower subband and excite them
to the upper subband (Fig.~\ref{impactionization}).

In our earlier work\cite{Kawamura2007}, we showed 
that dynamic nuclear polarization (DNP) is created in the breakdown regime 
of an odd-integer QHE.
We observed hysteretic voltage-current characteristic curves
and detected nuclear magnetic resonance by measuring the longitudinal voltage.
As observed in the other experiments on DNP pumping\cite{Kronmuller1999,Machida2002, Hashimoto2002,
Kane1992,Wald1994},
DNP is probably induced by the flip-flop process of electron spin $\mathbf{S}$
and nuclear spin $\mathbf{I}$ through the hyperfine interaction
${\cal H}_{\rm hyperfine} = A\mathbf{S} \cdot \mathbf{I}
 = A(S^{+}I^{-} + S^{-}I^{+})/2 + AS_{z}I_{z}$,
where $A$ is the hyperfine constant.
However, the detailed mechanism of  DNP in QHE breakdown regimes has not been understood yet.

In order to discuss the mechanism of DNP, 
at least the following two important aspects of DNP
should be clarified experimentally.
The first question that arises is where is DNP created,
and the second question that arises is whether  DNP is created only in the QHE breakdown regime.
It is natural to suppose that  DNP is created in the bulk part of the 2DEG
because  electric current mainly flows in the QH bulk channel 
under the breakdown regime.
On the other hand, however, it is well established that 
 the nuclear spins can be polarized by scattering of electrons between
the spin-resolved QH edge channels\cite{Machida2002,Kane1992,Wald1994}.
Therefore, it is crucial to distinguish which channel is relevant to the DNP
observed in the QHE breakdown regime.
It is also important to clarify
whether  DNP is created only in the breakdown regimes  or whether
it is created in the QH regimes as well.
If DNP is associated with the QHE breakdown, 
DNP should be created only when the bias electric field exceeds $E_{\rm c}$.
Thus, the bias electric field dependence of DNP gives 
a clue regarding the mechanism of  DNP.

In this paper, we report that DNP is created in a QH Corbino disk
where the edge channel transport is completely absent
and that DNP emerges 
near the critical voltage of the QHE breakdown.
The creation of DNP in the Corbino disk directly shows that 
DNP is created in the inner bulk part of the 2DEG 
and that the bulk channel current is relevant to DNP.
Furthermore, the agreement of the onset voltage of DNP 
with that of the QHE breakdown suggests that the underlying origin of DNP
is closely related to that of the QHE breakdown.

\begin{figure}[t]
	\begin{center}
		\includegraphics[width=5.5cm]{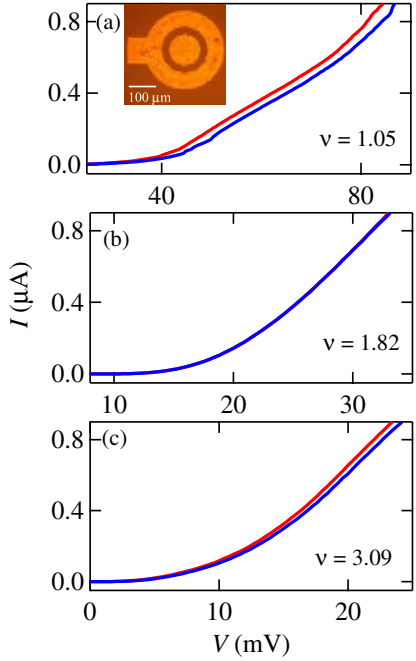}
		\caption{
		\label{IV}
			(Color online)
			$I$-$V$ curves obtained by sweeping the bias voltage 
			in the positive (blue) and negative (red) directions 
			at (a) $\nu$ = 1.05 ($B$ = 8.95 T),
			(b) $\nu$ = 1.82 ($B$ = 5.15 T), and (c) $\nu$ = 3.09 ($B$ = 3.04 T).
			Inset in (a): micrograph of the Corbino device.
		}
	\end{center}
	\end{figure}

Experiments were conducted using a sample with Corbino geometry
fabricated photolithographically
from a wafer of GaAs/Al$_{0.3}$Ga$_{0.7}$As single heterostructure
with 2DEG at the interface. 
The mobility and carrier density of the wafer are
$\mu$ = 115 m$^2$/Vs and $n$ =  2.27 $\times$ 10$^{15}$ m$^{-2}$, respectively.
The inner and  outer diameters of the Corbino disk
are 120 $\mu$m and 180 $\mu$m, respectively, as shown in the inset of
Fig.~\ref{IV}(a).
All the measurements were performed in a dilution 
refrigerator with a base temperature of 20 mK.
An external magnetic field $B$ was applied perpendicular to the 2DEG plane
by using a superconducting solenoid.
The current between the inner and outer electrodes
was measured using a standard dc method.
A single-turn coil around the device was used to irradiate
radio-frequency (rf) magnetic fields.

\begin{figure}[t]
	\begin{center}
		\includegraphics[width=8.5cm]{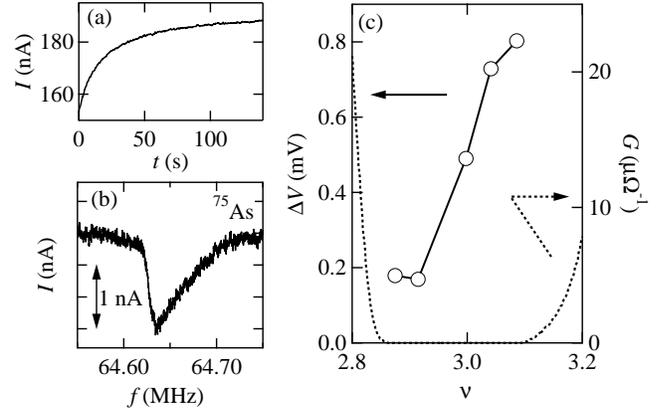}
		\caption{
		\label{NMR}
			(a) Time evolution of $I$ at $\nu$ = 1.05 
			after changing the bias voltage from $V$ = 0 mV to 50 mV at $t$ = 0.
			(b) NMR spectrum of $^{75}$As detected by measuring $I$.
			The frequency sweep rate is 13 kHz/min.
			(c) Filling factor dependence of the shift of $I$-$V$ curves $\Delta V$
			in the plateau region of $\nu$ = 3.
			Filling factor dependence of two-terminal conductance $G$ 
			is plotted by the dashed curve on the right axis.
			}
	\end{center}
	\end{figure}

Figures~\ref{IV}(a)-\ref{IV}(c) show
the current-voltage ($I$-$V$) curves 
at Landau level filling factors $\nu$ = 1.05 ($B$ = 8.95 T), $\nu$ = 1.82 ($B$= 5.15 T), and
$\nu$ = 3.09 ($B$ = 3.04 T), respectively. The curves are obtained by sweeping
the voltage in the positive (blue) and negative (red) directions
at a sweep rate of 0.67 mV/s.
Shifts of the $I$-$V$ curves are evident
in Figs.~\ref{IV}(a) and \ref{IV}(c), while no shift appears
in Fig.~\ref{IV}(b).
Shifts of  the $I$-$V$ curves are also evident at other filling factors 
in  the plateau regions of odd-integer QH states,
while they are not observed in the plateau regions of even-integer QH states.
Figure~\ref{NMR}(a) shows the time evolution of $I$ at 
$\nu$ = 1.05 after changing the voltage from $V$ = 0 mV to 50 mV at $t$ = 0.
The value of $I$ at $V$ = 50 mV increases slowly with a long relaxation time of  over 100 s,
which is typical to nuclear spin related 
phenomena\cite{Kronmuller1999,Machida2002,Hashimoto2002, Kane1992, Wald1994}.
The involvement of nuclear spins is clearly shown by 
the nuclear magnetic resonance measurement, as shown in Fig.~\ref{NMR}(b).
The current decreases at the resonant frequency of $^{75}$As.
The hysteretic $I$-$V$ curves, the slow time evolution, and the detection of NMR
show that DNP is created in the Corbino device.
Since the edge channel is absent in the Corbino disk,
the presence of  DNP directly shows
that nuclear spin polarization is created and detected in the bulk part of the QH conductor.
The shifts of the $I$-$V$ curves toward the smaller voltage sides indicate
that  DNP accelerates the QHE breakdown
by reducing the energy gaps of the odd-integer QH states.
Since the odd-integer QH states are stabilized by
the spin-splitting energy for electrons
$\Delta_{\rm s} = |g^{*}|\mu_{\rm B}B - A\langle I_z \rangle$,
where $g^{*}$ is the effective g-factor for electrons and $\mu_{\rm B}$
is the Bohr magneton,
the acceleration of the QHE breakdown
indicates that the nuclear spins are polarized upward ($\langle I_z \rangle >0$).

In Fig.~\ref{NMR}(c),
the shifts of the $I$-$V$ curves ($\Delta V$) defined at $I$ = 2 $\mu$A are
plotted against $\nu$ in the range of the QH plateau of $\nu$ = 3.
The amount of  shift increases with $\nu$.
This trend is similar to  that observed in  Hall-bar devices\cite{Kawamura2007}.
Although the origin of the trends is not understood,
the observation of similar filling-factor dependence in both Corbino and Hall-bar samples
suggests that the filling-factor dependence reflects the nature of the bulk channel.

To  understand how  DNP develops as a function of the bias voltage,
 we investigate the pumping voltage ($V_{\rm pump}$) dependence of  DNP.
If  DNP is induced by the avalanche multiplication 
of $e$-$h$ pairs\cite{Komiyama1996,Komiyama2000},
it is created only in the breakdown regime where
the pumping voltage $V_{\rm pump}$ exceeds the critical voltage of the QHE breakdown $V_{\rm c}$.
We employed the  following pump and probe method:
first, nuclear spin polarization is pumped by applying 
a pumping voltage $V_{\rm pump}$ for $t_{\rm pump}$ = 900 s.
Then, the voltage is suddenly changed to $V_{\rm probe}$ = 50 mV
and the value of $I$ is recorded.
Since  DNP is not created at $V_{\rm pump}$ = 0 mV,
the difference $\Delta I = I(V_{\rm pump}) - I(V_{\rm pump} =$ 0 ${\rm mV})$
originates from the DNP created at $V_{\rm pump}$.
Thus, $\Delta I$ functions as a measure of the amplitude of DNP.

\begin{figure}[t]
	\begin{center}
		\includegraphics[width=6cm]{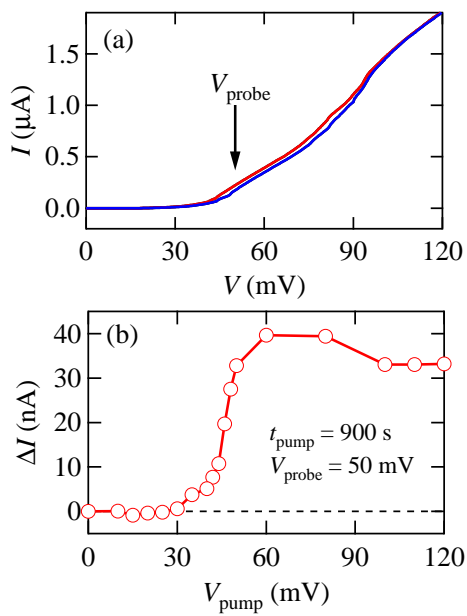}
		\caption{
		\label{PumpAndProbe}
			(Color online)
			(a) $I$-$V$ curves at $\nu$ = 1.05 ($B$ = 8.95 T)
			obtained by sweeping the voltage in the positive (blue) and negative (red) directions.
			(b) $V_{\rm pump}$ dependence of $\Delta I$ observed by the pump
				and probe method described in the text.
		}
	\end{center}
	\end{figure}

The $V_{\rm pump}$ dependence of $\Delta I$ at $\nu$ = 1.05
is measured as shown in Fig.~\ref{PumpAndProbe}(b).
In the range of $0 < V_{\rm pump} <$ 30 mV, the value of $\Delta I$
is almost zero.
The value of $\Delta I$ increases abruptly at about $V_{\rm pump}$ = 45 mV
and saturates above $V_{\rm pump}$ = 60 mV.
These results show that  DNP is not created 
in  QH states with vanishing longitudinal conductance
where $V_{\rm pump}$ is smaller than $V_{\rm c}$.
DNP emerges near $V_{\rm pump} = V_{\rm c}$
and saturates for larger $V_{\rm pump}$.
Similar results were  obtained in a Hall-bar sample used 
in Ref.~\citen{Kawamura2008}.
These  suggest that the underlying origin of DNP
is closely related to that of the QHE breakdown.

Since the QHE breakdown is induced by the avalanche-multiplication 
of the $e$-$h$ pairs due to inter-Landau-level impact ionization\cite{Komiyama1996,Komiyama2000},
it can be surmised that avalanche-multiplication is relevant to DNP.
In the case of  impact ionization in an odd-integer QH state,
electrons are excited from the up-spin subband to the down-spin subband,
 as indicated by the arrow {\it b}$\rightarrow${\it b}$^{\prime}$ in Fig.~\ref{impactionization}.
This process is accompanied by the up-to-down flip of electron spin.
The up-to-down flips of electron spins can cause the down-to-up flops of nuclear spins  
via the hyperfine interaction.
As a result, the nuclear spins are expected to be polarized upward ($\langle I_{z} \rangle > 0$),
thereby reducing the spin-splitting energy $\Delta_{\rm s}$.
This is consistent with the observation of the shifts of the $I$-$V$ curves
toward the smaller voltage sides [Figs.~\ref{IV}(a) and \ref{IV}(c)].
From these results, we infer that  DNP is induced by the spin-flip process of electrons 
in the inter-Landau-level impact ionization
during the avalanche multiplication of $e$-$h$ pairs.

To summarize, we have experimentally demonstrated that
the nuclear spins are polarized in the QH Corbino disk
under QHE breakdown regimes.
The pumping voltage
dependence of  DNP shows that  DNP starts to increase 
near the critical voltage for the QHE breakdown.
The presence of  DNP in the Corbino disk 
strongly suggests that the nuclear spin polarization in the bulk channel of the 2DEG
is pumped and detected.
The agreement of the onset voltage of  DNP with that of the QHE breakdown
suggests that the underlying origin of  DNP is  closely related to 
that of the QHE breakdown.
We suggest that DNP possibly originates due to  the electron spin flip process
in inter-Landau-level impact ionization
during the avalanche multiplication of $e$-$h$ pairs.

\section*{Acknowledgments}
This work is supported by a Grant-in-Aid from 
MEXT and the Special Coordination Funds for Promoting 
Science and Technology.


\begin{thebibliography}{99}
\bibitem{Ebert1983}
	G. Ebert, K. von Klitzing, K. Ploog, and G. Weimann:
	J. Phys. C {\bf 16} (1983) 5441.
\bibitem{Cage1983}
	M. E. Cage, R. F. Dziuba, B. F. Field, E. R. Williams,
	S. M. Girvin, A. C. Gossard, D. C. Tsui, and R. J. Wagner:
	Phys. Rev. Lett. {\bf 51} (1983) 1374.
\bibitem{Dolgopolov1992}
	V. T. Dolgopolov, A. A. Shashkin, N. B. Zhitenev,
	S. I. Dorozhkin, and K. von Klitzing:
	Phys. Rev. B {\bf 46} (1992) 12560.
\bibitem{Nachtwei1999}
	G. Nachtwei: Physica E {\bf 4} (1999) 79.
\bibitem{Komiyama1996}
	S. Komiyama, Y. Kawaguchi, T. Osada, and Y. Shiraki:
	Phys. Rev. Lett. {\bf 77} (1996) 558.
\bibitem{Komiyama2000}
	S. Komiyama and Y. Kawaguchi:
	Phys. Rev. B {\bf 61} (2000) 2014.
\bibitem{Kawamura2007}
	M. Kawamura, H. Takahashi, K. Sugihara, S. Masubuchi, K. Hamaya, 
	and T. Machida:
	Appl. Phys. Lett. {\bf 90} (2007) 022102.
\bibitem{Kronmuller1999}
	S. Kronm$\ddot{\rm u}$ller, W. Dietsche, K. von Klitzing,
	G. Denninger, W. Wegscheider, and M. Bichler:
	Phys. Rev. Lett. {\bf 82} (1999) 4070.
\bibitem{Machida2002}
	T. Machida, T. Yamazaki, K. Ikushima, and S. Komiyama:
	Appl. Phys. Lett. {\bf 82} (2003) 409.
\bibitem{Hashimoto2002}
	K. Hashimoto, K. Muraki, T. Saku, and Y. Hirayama:
	Phys. Rev. Lett. {\bf 88} (2002) 176601.
\bibitem{Kane1992}
	B. E. Kane, L. N. Pfeiffer, and K. W. West:
	Phys. Rev. B {\bf 46} (1992) 7264.
\bibitem{Wald1994}
	K. R. Wald, L. P. Kouwenhoven, P. L. McEuen, N. C. van der Vaart,
	and C. T. Foxon:
	Phys. Rev. Lett. {\bf 73} (1994) 1011.
\bibitem{Kawamura2008}
	M. Kawamura, H. Takahashi, S. Masubuchi, Y. Hashimoto, S. Katsumoto,
	K. Hamaya, and T. Machida:
	to be published in Physica E, arXiv:0711.3259.
\end{thebibliography}
\end{document}